\documentclass[preprint,showpacs,preprintnumbers,amsmath,amssymb]{revtex4}

\usepackage{graphicx}
\usepackage{dcolumn}
\usepackage{bm}

\begin{document}

\preprint{version:2.0-081208}

\title{Temperature Dependence of Critical Current Fluctuations in Nb/AlO$\mathrm{_{x}}$/Nb Josephson Junctions}

\author{Shawn Pottorf}
\author{Vijay Patel}
\author{James E. Lukens}
\affiliation{Department of Physics and Astronomy, Stony Brook University, Stony Brook, New York 11794--3800}
\email{shawn.pottorf@sunysb.edu}
\date{\today}

\begin{abstract}
We have measured the low frequency critical current noise in Nb/AlO$_{\mathrm{x}}$/Nb Josephson junctions. Unshunted junctions biased above the gap voltage and resistively shunted junctions biased near the critical current, $I_{c}$, have been measured. For both, the spectral density of $\delta I_{c}/I_{c}$, $S_{i_{c}}(f)$, is proportional to $1/f$, scales inversely as the area, $A$, and is independent of $J_{c} \equiv I_{c}/A$ over a factor of nearly 20 in $J_{c}$. For all devices measured at 4.2 K, $S_{i_{c}}$(1 Hz)$= 2.0 \pm 0.4 \cdot 10^{-12}$/Hz when scaled to $A=1$ $\mu$m$^{2}$. We find that, from 4.2 K to 0.46 K, $S_{i_{c}}(f)$ decreases linearly with temperature.
\end{abstract}

\pacs{74.40.+k, 74.78.Db, 74.78.Na, 85.25.Am, 85.25.Cp}

\maketitle



There has been a renewed interest in understanding the ubiquitous 1/$f$
fluctuations in the critical current and resistance in Josephson junctions.
This is motivated in part by the intense effort to reduce the noise-induced
decoherence in a variety of qubits based on these junctions. The dominant
model for these fluctuations is a collection of two level fluctuators (TLFs) in the oxide barriers of the junction that randomly and independently switch on and off the conductivity and supercurrent in small localized areas of the junction. This model \cite{rogers1985} gives rise to the expectation that the normalized spectral density of the critical current and resistance noise, e.g. $S_{i_{c}}\equiv S_{Ic}/I_{c}^{2}$ or $S_{r_{J}}\equiv
S_{R_{J}}/ R_{J}^{2}$, should be roughly independent of the junction
conductivity (or critical current density, $J_{c}$) and should scale
inversely as the junction area, $A$. It has been shown that the $I_{c}R_{J}$ product of the junction is unaffected by these fluctuations \cite{rogers1985}. Thus the two normalized spectra should be equal.

A survey of critical current fluctuations in Nb junctions from a number of
laboratories \cite{vanharlingen2004} recently found that this normalized
spectral density, scaled to an area $A=1$ $\mu $m$^{2}$, was indeed roughly
constant, had a value of $1.4\cdot 10^{-10}$/Hz at 4.2 K and varied with
temperatures as $T^{2}$. These results were then explained theoretically \cite{kenyon2000}. More recently, measurements of resistance fluctuations \cite{eroms2006} in submicron Al junctions made using shadow evaporation gave rather different results: the normalized and scaled spectral density was nearly three orders of magnitude smaller at 4.2 K, but varied linearly with temperature from about 100 K to nearly 0.8 K, where it appeared to saturate. It has been suggested \cite{faoro2007} that a possible explanation for this is that fluctuations in normal and superconducting junctions arise from different mechanisms. Very recently a prediction has appeared \cite{constantin2007}, based on the independent fluctuator model, that these fluctuations should vary linearly with $T$ but should increase as the fifth power of the thickness of the barrier, $t_{b}$. Since both $R_{J}$ and $I_{c}$ depend exponentially on $t_{b}$, this implies a small, but potentially detectable, dependence of the normalized spectral density of the fluctuations of these parameters on their values. In this Letter, we present data for the dependencies of these fluctuations on junction area, critical current density and temperature in Josephson junctions fabricated using a well-characterized, very versatile trilayer (Nb/AlO$_{\textrm{x}}$/Nb) process where the values of the barrier thickness as a function of $J_{c}$ have been accurately determined.


Examples of the normalized noise spectra we observe from our junctions are
shown in Fig. \ref{I-cfluct-fig1}. Here and throughout the paper, measured
noise spectra are presented in the terms of an equivalent current noise source, $S_{I}$ (or $S_{i}\equiv S_{I}/I_{c}^{2}$), shunting the junction. Spectra
shown are for three junctions with areas ranging from 3 $\mu$m$^{2}$ to 94 $\mu$m$^{2}$ measured at 4.2K. These junctions, having $J_{c}=100$ A/cm$^{2}$, have no external shunts and are biased at 3.7 mV, which is above the gap voltage. The data were obtained by placing the junction as one arm of a Wheatstone bridge having a commercial SQUID as a null detector. In each case the spectra are fit to the sum of white noise plus a $1/f$ spectrum, $S_{1/f}$. The white noise component is just that expected from the shot noise due to the bias current, which dominates other white noise sources, e.g. the Johnson noise due to the other arms of the bridge circuit. $S_{i_{c}}$ is obtained from the $1/f$ component of the spectra corrected for the bias conditions, e.g. for unshunted junctions
biased above the gap voltage the $1/f$ noise should almost be entirely
due to resistance fluctuations. For this case, the constancy of the $I_{c}R_{J}$ product gives $S_{i_{c}}=S_{1/f}\frac{I_{c}^{2}}{I_{b}^{2}}$.

Throughout the paper, the values presented for $S_{1/f}$ and related quantities are at 1 Hz. The inset of Fig. \ref{I-cfluct-fig1} shows $S_{1/f}$ for the three junctions as a function of $A$, along with a line having the predicted linear dependence with $A$. This clearly demonstrates the scaling expected from the independent fluctuator model. To facilitate the comparison of $S_{i_{c}}$ of a variety of junctions, we present data for $S_{i_{c}}$ scaled to $A=1$ $\mu $m$^{2}$ and denoted by $S_{i_{c}}^{s}.$ For the junctions in Fig. \ref{I-cfluct-fig1}, $S_{i_{c}}^{s}=2.2\cdot 10^{-12}$/Hz, which is about two orders of magnitude less than that commonly seen in other superconducting junctions \cite{vanharlingen2004}. We have further verified that the noise is consistent with resistance fluctuations by observing that $S_{1/f}$ is proportional to $I_{b}^{2}$ between 3.7 mV and 6 mV, i.e. $S_{R_{J}}$ is independent of $I_{b}$, where $I_{b}$ is the total dc current through the junction. Such unshunted junctions are not suitable for a study of the temperature dependence of this noise due to the high power dissipation at the required bias voltage that leads to substantial electron and lattice heating at lower temperatures \cite{wellstood1994}. Therefore, to extend this work to lower temperatures we fabricated shunt resistors in parallel with the junction to form a non-hysteretic resistively shunted junction (RSJ).


All junctions reported here were fabricated in our laboratory using a process
(SAL-EBL) that has been detailed previously \cite{patel2005}. These junction
have been shown to be of very high quality and dimensional uniformity. This high quality is also demonstrated by very small subgap leakage currents, e.g. a subgap resistance, measured at 400 mK and 0.5 mV that is over $10^{6}R_{J}$ \cite{patel2005}. The initial step in the fabrication of the Nb/AlO$_{\textrm{x}}$/Nb junctions is the \textit{in situ} sequential deposition of the trilayer through a liftoff resist mask. The tunnel barrier is formed by thermal oxidation of the 80 nm Al interlayer in dry O$_{2}$ for exposures of 10$^{1}-10^{4}$ Torr-min (10$^{5}-10^{8}$ Pa-s) giving the range of $J_{c}$ of $50-800$ A/cm$^{2}$ studied here. The properties of the barriers, i.e. barrier height and thickness, have been carefully determined by fitting their normal conductance vs. voltage using numerical solutions of the Schr\"{o}dinger equation \cite{tolpygo2003}.

For RSJs, the shunt resistors (see Fig. \ref{I-cfluct-fig2}, inset) were
designed to minimize their parasitic inductance, e.g. by placing a Nb ground
plane under the resistor, while having a sufficient volume and contact area
to ensure adequate cooling of both the electrons and the lattice. To ensure
nonhysteretic I-V curves, the resistance of the shunt resistor was selected to give $\beta _{c}\simeq 0.3$, where $\beta _{c} \equiv 4\pi e I_{c}R^{2}C/h$. With this constraint the ratio of the 1/$f$ noise to the Johnson noise due to the resistor should scale as $J_{c}^{3/2}$. This fact, along with rapidly increasing oxidation times, sets a practical lower limit on $J_{c}$. The upper limit is set by a degradation of the barrier quality, as seen by increasing subgap leakage, for $J_{c}>5$ kA/cm$^{2}$ \cite{kleinsasser1993,patel2005}.

The temperature rise, $\delta T$, due to the power, $P$, dissipated in the
resistor can be described by $\delta T=(kP+T_{0}^{n})^{1/n} - T_{0}$, where k is a constant depending on the material and geometry and $T_{0}$ is the substrate temperature. We have measured this heating in resistors having the same design as our shunt resistors by observing the increase in resistor temperature (measured using noise thermometry) as a function of $P$ over the range of temperatures used. These data are shown in Fig. \ref{I-cfluct-fig2} for temperatures ranging from 0.35 K through 0.7 K. As one can see, the assumed fitting functions with $n=4.1 \pm 0.1$, which is consistent with lattice heating \cite{wellstood1994}, and $k=22.5 \cdot 10^{-3} \pm 0.9$ (K$^{n}/$nW) provide good descriptions of the temperature increase, and can be used to provide heating corrections to the data if needed. The noise spectra for the RSJs were generally measured at an input power of about 1 nW. For this low power, heating makes a significant, though small, correction to the temperature of the RSJs at the lowest temperatures in the measurement.


The junctions were cooled using a $^{3}$He refrigerator having a base
temperature of 350 mK. Filtering at room temperature and 4.2 K along with
battery powered amplifiers and bias current sources were used to isolate the
junctions from electronic noise. This filtering, along with the bridge
measurement circuit reduced the bias current noise to levels too low to be
detected. The upper arms of the bridge are $\sim 1$ k$\Omega$ resistors anchored at 4.2 K. The lower arms consist of the sample along with a matched balancing resistor, $R_{bal}$. This balancing resistor was either a metal film resistor attached to the chip carrier in the case of unshunted junctions, or, for RSJs, an on-chip resistor of the same design as the junction shunt resistor, $R_{sh}$ (see Fig. 2 inset). The SQUID null detector had a noise limit of 0.5 pA$^{2}$/Hz above 10 Hz and a $1/f$ knee at 1 Hz when directly connected to the bridge.

The noise spectra from the RSJs could also be fit to the sum of white
noise plus $1/f$ noise. In general, the values of these two quantities were
anomalous at higher voltages where clear resonances were observed. These behaviors can be seen in the inset of Fig. \ref{I-cfluct-fig3}, which shows the dynamic resistance along with $S_{i_{c}}^{s}$ and $S_{i}^{s}$(1 kHz). These quantities are plotted vs $\Delta i\equiv (I_{b}-I_{c})/I_{c}$ on a log scale so the behavior near $I_{c}$ is clearly visible. The resonances were due in part to the parasitic inductance ($L_{p}=2.3$ pH) of the shunt resistor. This behavior made it impossible to use the value of the white noise as a reliable thermometer. However, the values of $S_{i_{c}}$ extracted from the spectra are independent of bias at low voltages, i.e. for bias currents just above $I_{c}$. In this bias-independent region, the values of $S_{i_{c}}$ obtained also agrees with those obtained from unshunted junctions on the same wafer at 4.2 K where heating in the unshunted junctions is not an issue. Figure \ref{I-cfluct-fig3} compares the values of $S_{i_{c}}^{s}$ obtained at 4.2 K for both shunted and unshunted junctions with areas in the range of 3 $\mu$m$^{2}$ to 94 $\mu$m$^{2}$ and $J_{c}$ values from 55 to 807 A/cm$^{2}$. Within the scatter of about a factor of 2, $S_{i_{c}}^{s}$ is independent of area and $J_{c}$ as expected from the simple independent fluctuator model. The temperature dependence of $S_{i_{c}}^{s}$ for our RSJs measured near $I_{c}$ is shown in Fig. \ref{I-cfluct-fig4}. A best fit (dashed line) having a linear temperature dependence clearly describes the data quite well. At 4.2 K $S_{i_{c}}^{s}=1.42 \cdot 10^{-12}/$Hz and decreases to $0.16 \cdot 10^{-12}/$Hz at the lowest temperature, 0.46 K. The lowest temperature data points were corrected for heating in the shunt resistor at the measurement power of about 1 nW. The largest correction at the lowest temperature was $\sim$90 mK, but the conclusion that $S_{i_{c}}^{s}$ varies linearly with $T$ would not be significantly changed even if this correction were omitted.


A recent analysis of critical current noise due to TLFs in the junction barrier \cite{constantin2007} predicts that this noise is given by $S_{i_{c}}^s \sim (T/f)t_{b}^{5}$, i.e. it increases as the fifth power of the junction barrier thickness, $t_{b}$. The solid line in Fig. \ref{I-cfluct-fig3} shows the best fit to this equation using the proportionality constant as the fitting parameter. This proportionality constant contains the density of states of the TLFs, which is therefore assumed to be independent of $J_{c}$. We have previously done careful comparisons of the dependencies of both $J_{c}$ and $t_{b}$ on oxidation parameters \cite{tolpygo2003}, which allows us to accurately determine the barrier thickness corresponding to a given $J_{c}$. While the predicted variation of $S_{i_{c}}^{s}$ on $t_{b}$ is not seen, we must note that there is no independent measure of the density of TLFs. There is some reason to suspect that this density is increasing with $J_{c}$, since for $J_{c} > 5$ kA/cm$^{2}$ it is well known that barrier leakage increases significantly \cite{kleinsasser1993}.

In summary, we have measured $1/f$ fluctuations in the critical current and resistance of Josephson junctions and observed their dependence on junction area, critical current density and temperature. We find that $S_{I_{c}}$ scales linearly with temperature, without saturation, down to the lowest temperature measured (0.46 K). In addition, $S_{i_{c}}^{s}$, averaged over all junctions measured, gives $2.0 \pm 0.4 \cdot 10^{-12}$/Hz at 1 Hz and 4.2 K, which is nearly two orders of magnitude less than that commonly observed in Josephson junctions. However, the data do not exhibit the recently predicted \cite{constantin2007} dependence of $S_{i_{c}}^{s}$ on barrier thickness as $t_{b}^{5}$.

This work was supported in part by NSF grant \#DMR-0325551.


\pagebreak

\pagebreak

\begin{figure}
\centering
\caption{Current noise spectra, $S_{I}$, of unshunted junctions having areas of 3 $\mu$m$^{2}$, 22 $\mu$m$^{2}$ and 94 $\mu$m$^{2}$ at 4.2 K and $J_{c}=100$ A/cm$^{2}$. The junctions were biased at 3.7 mV. The solid curves through the spectra are best fits to white plus $1/f$ noise as described in the text. The inset shows the 1/$f$ component of the noise spectra at 1 Hz as a function of junction area with a linear fit to the data (red line).}
\label{I-cfluct-fig1}
\end{figure}

\begin{figure}
\centering
\caption{Temperature increase as a function of power dissipated in
AuPd thin film resistors for bath temperatures of 0.35 K to 0.7 K. The length ($L$), width ($W$) and thickness ($t$) of each resistor are $L=33.5$ $\mu$m, $W=28$ $\mu$m, $t=0.08$ $\mu$m. The inset shows the Wheatstone bridge circuit with a thin film AuPd resistor and an RSJ in the lower half of the bridge circuit (note that the RSJ was replaced by a sample resistor for these data).}
\label{I-cfluct-fig2}
\end{figure}

\begin{figure}[!]
\caption{$S_{i_{c}}^{s}$, at 4.2 K of shunted (solid black symbols) and unshunted (open blue symbols) junctions as a function of $J_{c}$ for nominal junction areas of 2.25 $\mu$m$^{2}$ (triangles), 3 $\mu$m$^{2}$ (circles), 22 $\mu$m$^{2}$ (diamonds), 94 $\mu$m$^{2}$ (squares). The solid red line is a best fit to the data from ref. \cite{constantin2007} as described in the text, and the dashed line is the value from ref. \cite{vanharlingen2004}. The inset shows $S_{i_{c}}^{s}$ (solid black circles) and the normalized noise at 1 kHz, $S_{i}^{s}$(1 kHz), (open green triangles) versus $\Delta i$ for an RSJ having $J_{c}=807$ A/cm$^{2}$ and $A=3.3$ $\mu$m$^{2}$. The green dashed line in the inset is the expected value of $S_{i}^{s}$(1 kHz).}
\label{I-cfluct-fig3}
\end{figure}

\begin{figure}
\caption{Temperature scaling of normalized critical current spectral density
at 1 Hz for two 3.3 $\protect\mu $m$^{2}$ RSJs from two different process runs represented by closed and open symbols. The RSJs had $I_{c}=27$ $\mu$A and $R_{sh}=4.4$ $\Omega $. The dashed line shows the best fit to the data having a linear dependence on $T$.}
\label{I-cfluct-fig4}
\end{figure}

\pagebreak

\begin{figure}
\centering
\includegraphics[width=6.5in]{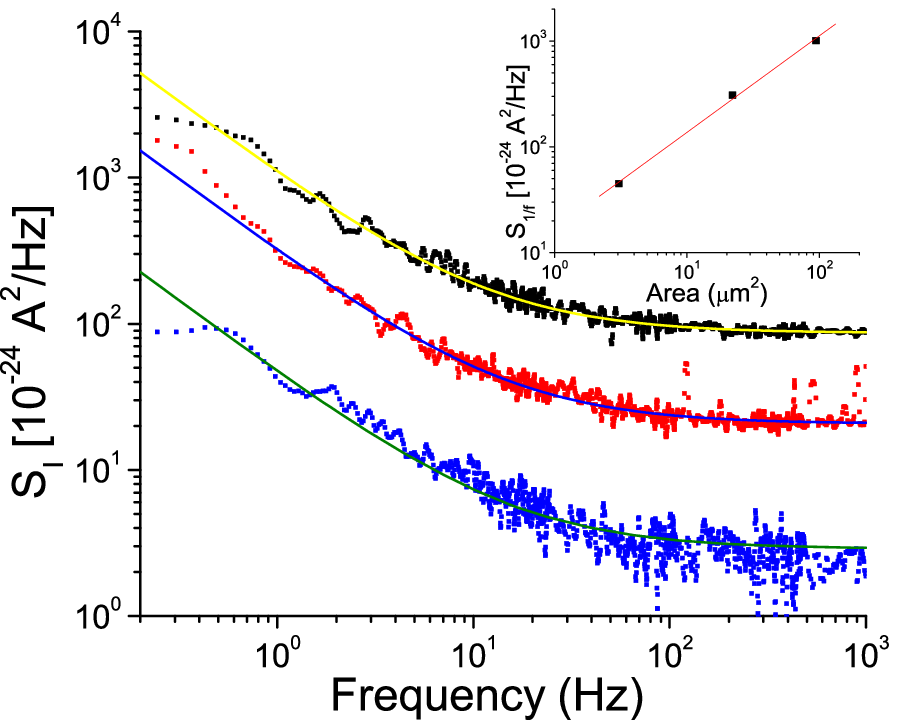}
\end{figure}

\pagebreak
\begin{figure}
\centering
\includegraphics[width=6.5in]{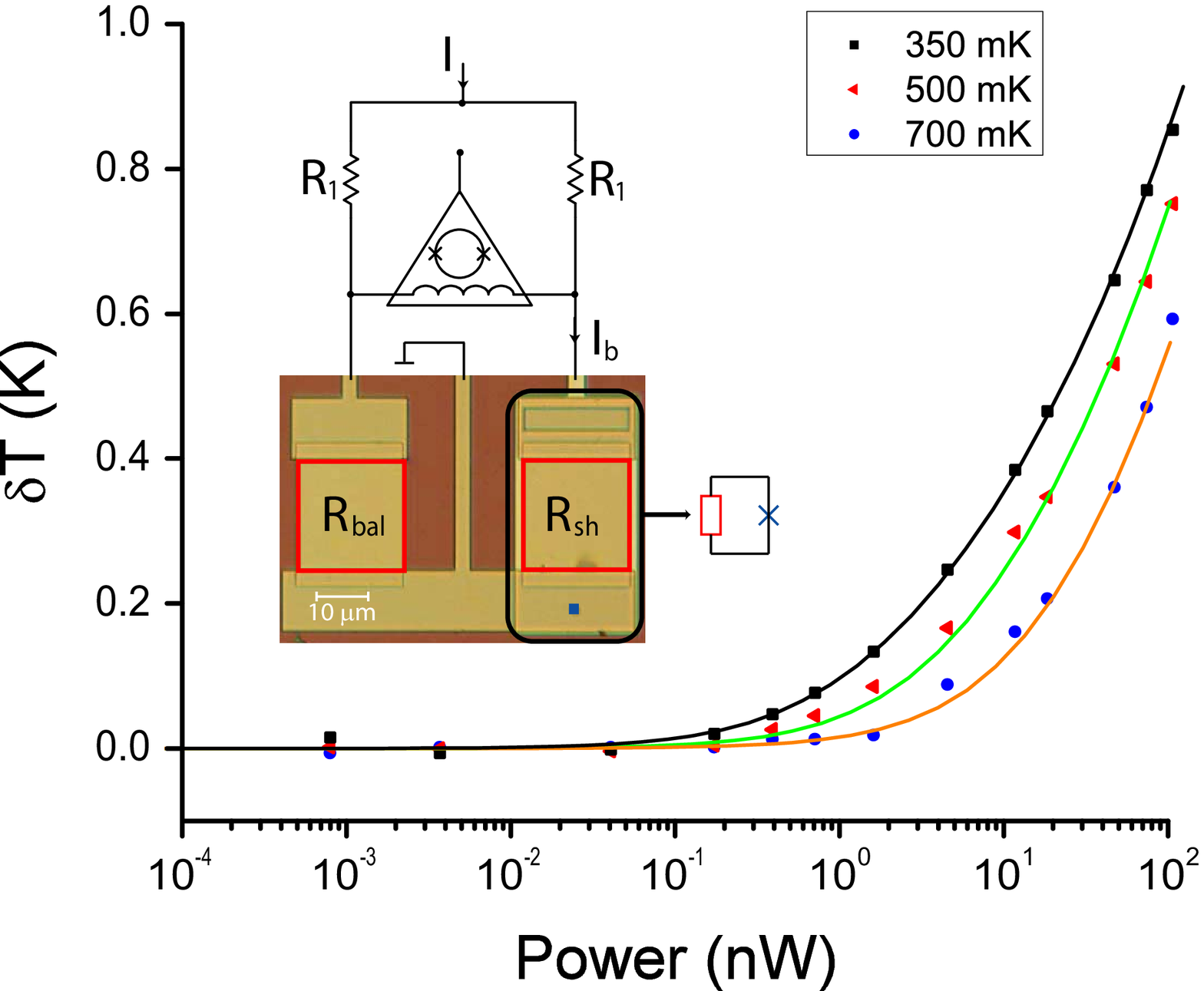}
\end{figure}

\pagebreak
\begin{figure}
\centering
\includegraphics[width=6.5in]{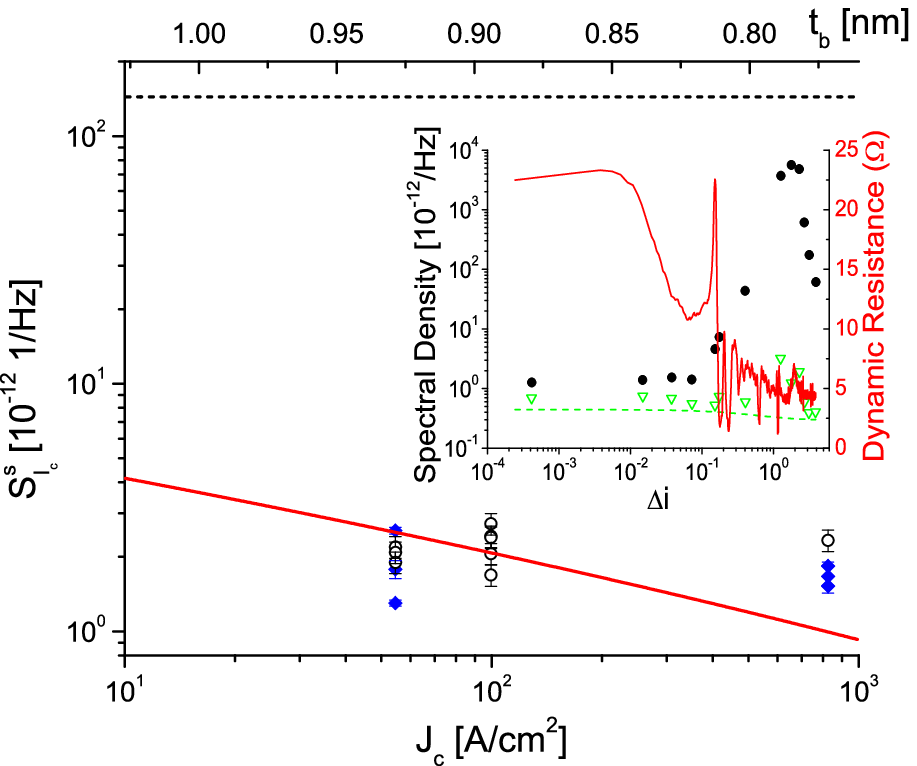}
\end{figure}

\pagebreak
\begin{figure}
\centering
\includegraphics[width=6.5in]{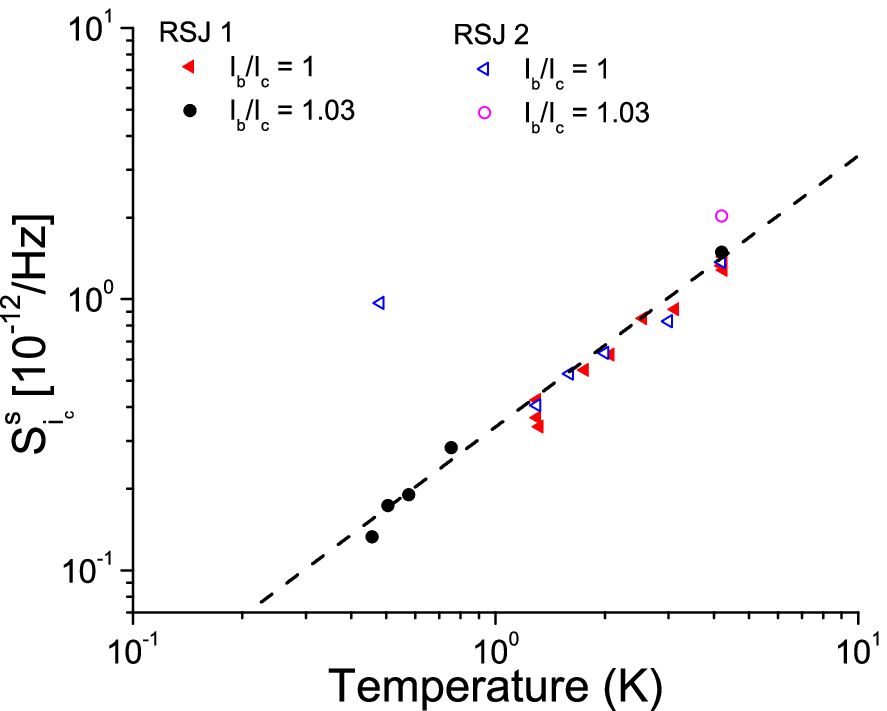}
\end{figure}

\end{document}